\documentclass[letterpaper,10pt]{article}

  \usepackage{osameet3}
  \usepackage{bm}

  \usepackage{cite}
  \usepackage{amsfonts}

  \DeclareGraphicsExtensions{.pdf,.png}

  \usepackage[cmex10]{amsmath}
  \usepackage{mathtools}
  \newcommand{\beq}{\begin{IEEEeqnarray}{rCl}}
  \newcommand{\eeq}{\end{IEEEeqnarray}}

  \usepackage[nomargin,inline,final]{fixme}
  \fxusetheme{color}
  \usepackage{hyperref}
  \hypersetup{
    unicode=false,
    pdftoolbar=true,
    pdfmenubar=true,
    pdffitwindow=false,
    pdfstartview={FitH},
    pdfcreator={pdflatex},
    pdfproducer={Latex with hyperref},
    pdfnewwindow=true,
    colorlinks=true,
    linkcolor=red,
    citecolor=green,
    filecolor=magenta,
    urlcolor=blue}

\begin{document}

%%% Preamble
  \title{Microresonator-enhanced, Waveguide-coupled Emission from Silicon Defect Centers for Superconducting Optoelectronic Networks}

  \author{A.~N.~Tait$^{*}$,
          S.~M.~Buckley,
          A.~N.~McCaughan,
          J.~T.~Chiles,\\
          S.~Nam,
          R.~P.~Mirin
          and J.~M.~Shainline,
          \\ \small National Institute of Standards and Technology, Boulder, CO 80305, USA
          \\ $^*$alexander.tait@nist.gov
          }

  % \maketitle

\begin{abstract}
  \noindent Superconducting optoelectronic networks could achieve scales unmatched in hardware-based neuromorphic computing. After summarizing recent progress in this area, we report new results in cryogenic silicon photonic light sources, components central to these architectures. \fxwarning{get rid of copyright}
\end{abstract}

% the review part
\section{Superconducting optoelectronic neural networks}
  % Something about neuromorphic
  \fxwarning{More intro/motivation before diving into SOENs?}
  Artificial neural networks are capable of a wide variety of tasks in control, simulation, and recognition. Specialized CMOS hardware has been developed to vastly reduce the power consumption of neuromorphic computations as compared to conventional, centralized architectures. CMOS neuromorphic hardware can outperform in certain application areas, but they are then primarily limited by interconnect performance. Extending the application areas of hardware neuromorphics will require new physical platforms, such as those based on photonics, analog electronics, or Josephson junctions. Superconducting optoelectronic neural networks (SOENs) offer a route to hardware neuromorphic systems with the largest scales considered to date~\cite{Shainline:17}. In the SOEN architecture, light is used for communication in order to avoid bandwidth/distance/dissipation tradeoffs affecting both analog and digital electronic interconnects. Signaling at the lowest possible light levels is achieved using single-photon detectors. Since these detectors must operate at cryogenic temperatures, other unusual physical phenomena can be employed. In particular, at these temperatures, crystal defect centers in silicon can emit light. W centers are ideal candidates because they emit at 1220~nm, a wavelength that can be guided by silicon-on-insulator (SOI) waveguides, and because they can be fabricated with a simple ion implantation. Cryogenic silicon photonic platforms to support SOENs are nascent but rapidly developing.

  \begin{figure}[b]
    \centering
    \includegraphics[width=.99\linewidth]{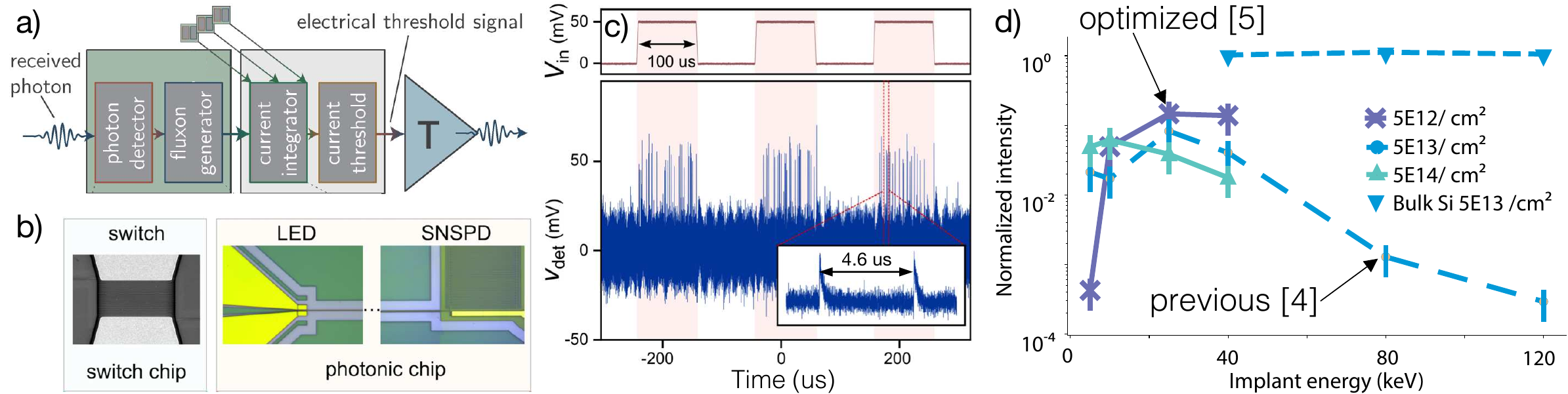}
    \caption{Recent results on Superconducting Optoelectronic Networks (SOENs). a) Latest conception of the SOEN neuron, adapted from~\cite{Shainline:19}. ``T'' is the optical transmitter consisting of a silicon LED. b) Part of a SOEN neuron: a thresholding, amplification, transmission, and detection drive train, adapted from~\cite{McCaughan:19}. c) Demonstration of the drive chain in (b). d) Optimization of photoluminescent brightness of W-center emitters in silicon-on-insulator substrates, adapted from~\cite{Buckley:19arxiv}.}
    \label{fig:recent_results}
  \end{figure}

  Fig.~\ref{fig:recent_results} represents selected results from the past 6 months of SOEN research. The latest conception of the neuron signal pathway~\cite{Shainline:19} is shown in Fig.~\ref{fig:recent_results}(a). Incoming photons are detected by superconducting nanowire single-photon detectors (SNSPDs). Induced currents are then summed and integrated in a magnetic flux loop. The resulting singal is thresholded and amplified to ultimately drive a W-center-based silicon light emitting diode (LED): the transmitter (T). A key part of this drive train was demonstrated in the circuit of Fig.~\ref{fig:recent_results}(b) from~\cite{McCaughan:19}. The circuit consisted of a high-impedance superconducting switch driving a silicon LED whose output is, in turn, detected in an SNSPD. The essential role of the switch is to transduce a superconductor-compatible (50~mV) pulse into a semiconductor-compatible (1~V) pulse that then drives the LED. The LED-to-SNSPD optical link was first demonstrated in~\cite{Buckley:17}. Fig.~\ref{fig:recent_results}(d) shows that, when 50~mV is applied to the switch, photon clicks are detected at the end of the link with a mean interval of 4.6~$\mu$s.

  The low heat dissipation density promised by SOENs depends strongly on a mastery of silicon photonic light sources: increasing brightness, radiative efficiency, waveguide coupling, etc. The properties of silicon emissive defects in photonics-compatible SOI have not been studied or optimized until recently. Figure~\ref{fig:recent_results}(d) shows an optimization of implant parameters performed in~\cite{Buckley:19arxiv}. Unlike in bulk Si (blue triangles), W-center brightness in SOI is strongly dependent on implant energy. Compared to prior work on waveguide-coupling, which used an 80~keV implant~\cite{Buckley:17,McCaughan:19}, photoluminescence (PL) brightness has been increased by two orders-of-magnitude.

\section{Silicon photoluminescence coupled to a waveguide and microdisk resonator} \label{sec:coupled-PL}
  In this work, we demonstrate coupling of optimized W centers to a silicon photonic circuit. The circuit -- consisting of a microdisk resonator, waveguide, and grating coupler -- enables spatial differentiation of free-space and waveguide-coupled emission. We show that waveguide-coupled emission is significantly enhanced at wavelengths corresponding to the modes of the resonator.
  Substantial research has been devoted to SOI-compatible emitters using III-V semiconductors, germanium, and rare-earths; however these each complicate fabrication with, respectively, epitaxy or wafer bonding, quantum wells, and Al\textsubscript{2}O\textsubscript{3} optics.
  % ~\cite{Roelkens:10} ~\cite{Pollnau:18}.
  W centers are created simply by damaging the silicon crystal though silicon ion bombardment and then annealing -- both of these steps are CMOS-compatible at the wafer scale. Embedding the emitters directly in the crystal presents a great simplification for applications below 40~K.
  W-center emission into a silicon-on-insulator waveguide has been observed before in an electrically injected LED~\cite{Buckley:17}. While electrical pumping is the eventual direction of our research, the optically pumped setup presents several advantages for research: 1) many devices can be probed in an array, meaning that the physics of the emitters can be better studied and device parameters better optimized; 2) photoluminescent properties can be studied independent of electrical factors (e.g. ohmic heating); 3) fabrication is 2-step (Si rib, Si$^+$~implant) instead of 7-step (Si rib, Si pedestal, Si$^+$~implant, heavy/light boron, heavy/light phosphorus). Coupling between W centers and resonator modes has not been shown before, although copper-based centers have been coupled to resonances of a suspended photonic crystal in~\cite{Sumikura:14}. Coupling resonance-enhanced silicon emission to a photonic integrated circuit has not been shown with any defect center.

  % \subsection{Fabrication}
    The substrate used is a 76.2~mm SOI wafer with 220~nm silicon device layer on 2~$\mu$m buried oxide. The wafer is patterned and implanted with $^{28}$Si$^+$ at an energy of 25~keV and fluence of 5$\times 10^{12}$/cm$^2$. The first mask is stripped. The wafer is patterned again and fully etched through the silicon device layer. Finally, the whole wafer is encapulated by a 1.5~$\mu$m PECVD oxide. More detail on fabrication can be found in~\cite{Buckley:19arxiv}. Waveguides are designed to be 350~nm wide such that they are single-mode at 1220~nm. Cutback tests indicate a single-mode waveguide loss of 18.9~dB/cm, which is significantly higher than other runs, we expect due to a new etch recipe.

    \begin{figure}[htb]
      \centering
      \includegraphics[width=.99\linewidth]{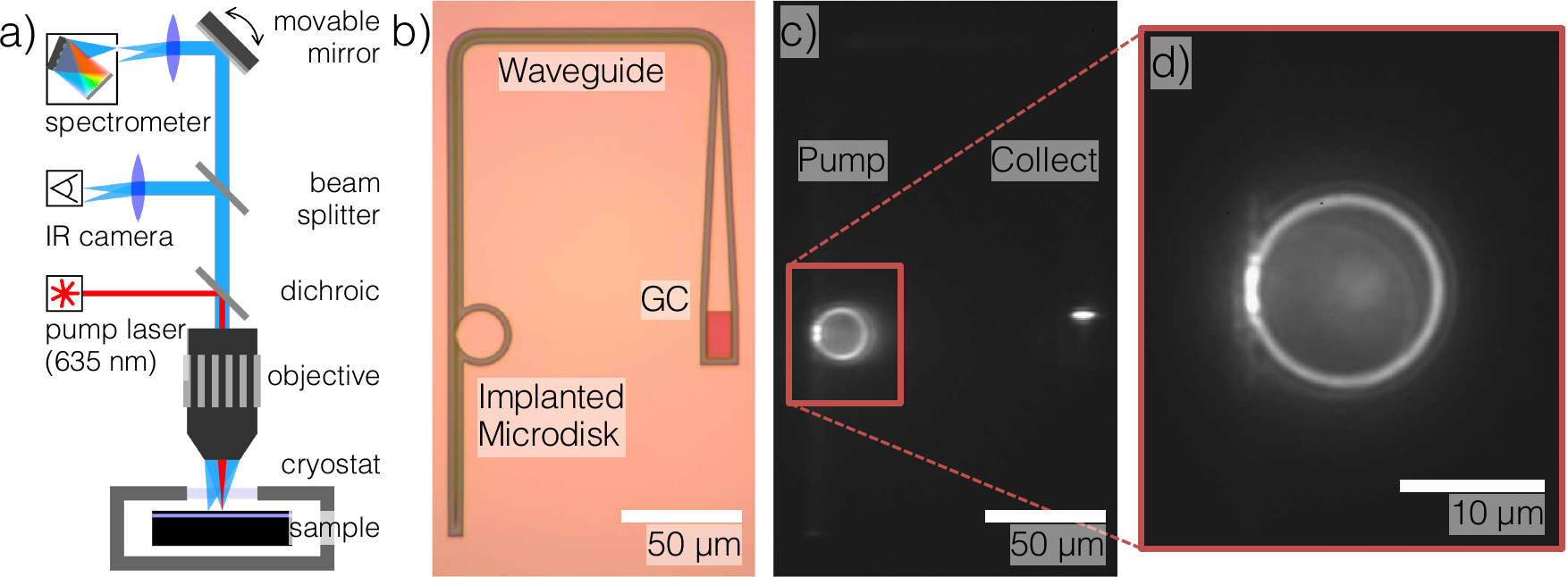}
      \caption{Setup and circuit. a) Simplified experimental setup for distinguishing waveguide-coupled photoluminescence (PL). b) Visible image of the circuit. c) Infrared image of PL from the circuit when the microdisk is pumped. Some PL exits directly into free-space. Other PL is waveguide-coupled to the GC. d) Zoomed in IR image of the microdisk.}
      \label{fig:setup_and_sample}
    \end{figure}

  % \subsection{Experimental setup} \label{sec:setup}
    We introduce an offset collection technique that allows for observation of waveguide-coupled PL. In the silicon photonic circuit, shown in Fig.~\ref{fig:setup_and_sample}(b), a microdisk resonator with 10~$\mu$m radius is implanted with W centers. It is coupled to a waveguide with a coupling gap of 100~nm. The waveguide is routed to a grating coupler (GC), designed to emit normal to the sample. When the microdisk is pumped, it emits upward into free-space with a spectrum similar to that of unpatterned SOI. Waveguide-coupled light, on the other hand, is routed to the GC. Since this GC is offset from the free-space emission source, the waveguide-coupled component is spatially discernable.

    A simplified experimental setup is shown in Fig.~\ref{fig:setup_and_sample}(a). An 11~mW laser pump at 635~nm (red) is focused onto a device with W centers. The long working distance objective has a numerical aperture of 0.42. Radiative recombination in W centers exceeds non-radiative pathways only below 45~K, so the sample is placed in an optical cryostat. Infrared PL around 1218~nm (depicted as blue) returns from the pumped microdisk itself and from the offset GC. When the returning light is imaged with a lens, the two emission sources are spatially separated upon reaching a slit entering the spectrometer. Rotating a mirror before the spectrometer allows selection between centered and offset collection.
    We verify this spatial resolving technique with an InGaAs camera, images from which are shown in Fig.~\ref{fig:setup_and_sample}(c, d).
    % (Princeton Instruments NIRvana 640)
    The perimeter of the disk appears brighter due to scattering of the microdisk modes off of the sidewalls. Additional scattering occurs at the coupling region leading to a brighter spot on the left of the disk.
    % Note that the outermost ring is not a double image -- the silicon slab is not etched \fxwarning{30 um} around the waveguides, meaning that three disks are seen: white is the microdisk sidewall; gray is some PL coupling through the buried oxide; and black is the silicon slab.

    \begin{figure}[htb]
      \centering
      \includegraphics[width=.7\linewidth]{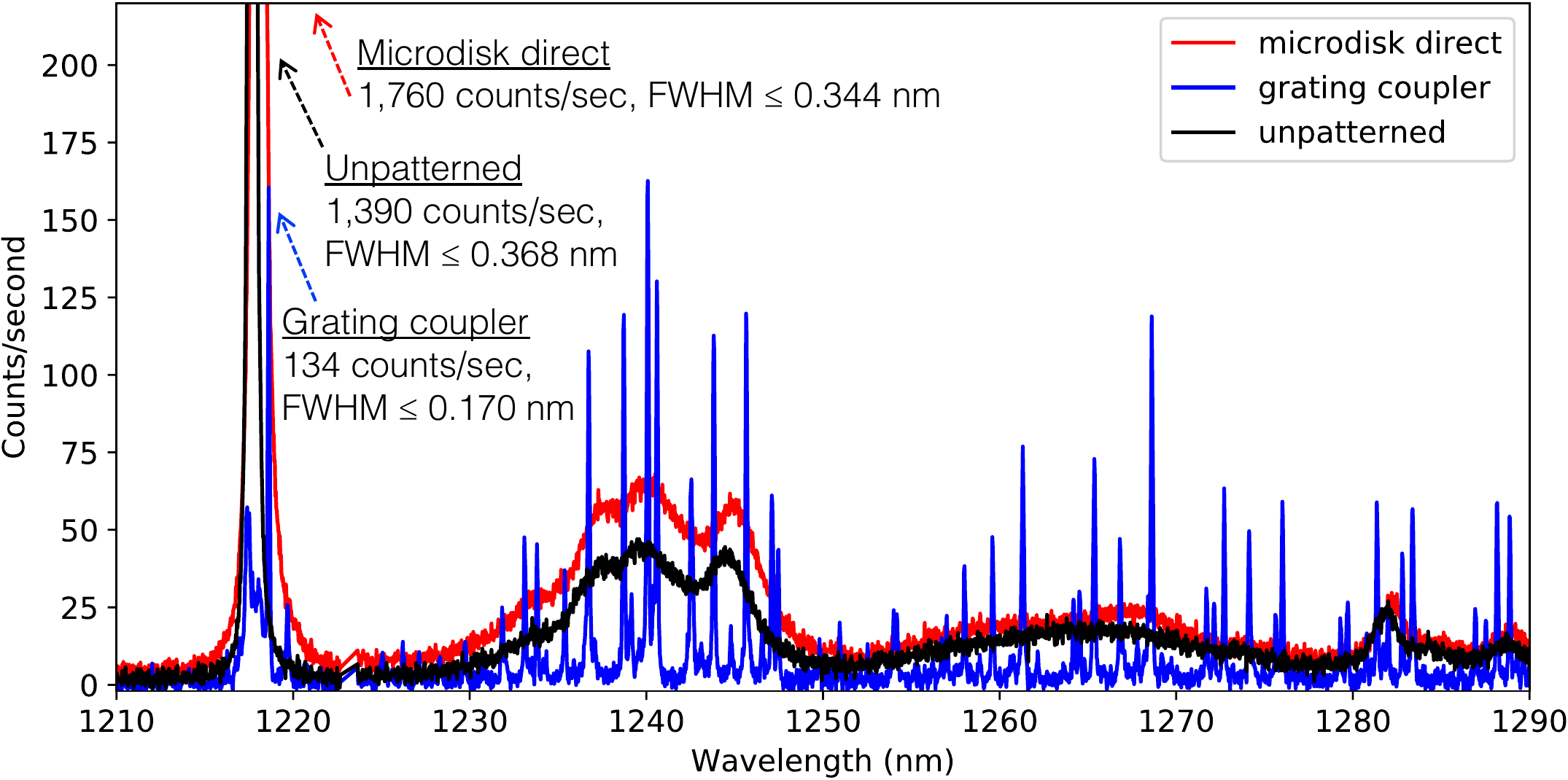}
      \caption{Comparison of spectra when collecting from directly over the top of the pumped microdisk (red), the offset GC (blue), and an unpatterned control region (black). The zero-phonon line amplitude and upper bound of full-width half maximum (FWHM) are indicated.}
      \label{fig:results}
    \end{figure}

  % \subsection{Results}
    Results indicate profound differences in the PL spectra of direct collection from the microdisk center (i.e., free-space coupled) and offset collection from the GC (i.e. microresonator- and waveguide-coupled). Fig.~\ref{fig:results} compares these with a control that is a large unpatterned, implanted region. The zero-phonon line (ZPL) at 1218~nm is brighter over the center of the microdisk than unpatterned due to scattering off of the device sidewalls. Otherwise the microdisk and control are similar. Collecting from the GC yields several high-Q peaks corresponding to the modes of the microdisk. That means waveguide-coupled emission is enhanced on resonance, even compared to the brighter controls. The ZPL from the GC is weak because the cavity is not on resonance. The quality factor of some peaks was measured up to 7,160, which is a lower bound as limited by the resolution of the spectrometer. Amplitudes are affected by the efficiency of the GC, but this was not characterized at the temperatures used. Within the same cooldown, stadard deviation is estimated to be 10\%, like in~\cite{Buckley:19arxiv}.

% \section{Conclusion}
  The demonstration of coupling W-center emission to a silicon photonic circuit represents progress in light sources for silicon photonics. Due to their ease of fabrication, they could play a significant role in large-scale photonic systems. Further work could also investigate Purcell enhancement, nonlinear phenomena, optical amplifiers, or electrical pumping.
  % Lasers offer light that is coherent and higher power. These properties are particularly favorable for inducing nonlinear effects, for example, photon pair generation through degenerate four wave mixing~\cite{Gentry:15}. Furthermore, lasers can exhibit dynamics: excitability has been shown to emulate spiking neuron behavior~\cite{Shastri:16}, and Q-switching and mode-locking can substantially increase instantaneous power.

  \begin{center}
    This is a contribution of NIST, an agency of the U.S. government, not subject to copyright.
  \end{center}

% \vfill
\bibliographystyle{osajnl}
% \bibliography{Master_Biblio_NIST}  % From master, full records
% Then run ./bibscript
% \bibliography{ofc-emission}  % From local, full records
% \bibliography{ofc-emission_proc}  % Short versions

\begin{thebibliography}{1}
\newcommand{\enquote}[1]{``#1''}

\bibitem{Shainline:17}
J.~M. Shainline \emph{et~al.}, Phys. Rev. Applied \textbf{7}, 034013 (2017).

\bibitem{Shainline:19}
J.~M. Shainline \emph{et~al.}, Journal of Applied Physics \textbf{126}, 044902
  (2019).

\bibitem{McCaughan:19}
A.~N. McCaughan \emph{et~al.}, Nature Electronics \textbf{2}, 451--456 (2019).

\bibitem{Buckley:19arxiv}
S.~M. Buckley \emph{et~al.}, arXiv eprint, arXiv:1911.01317  (2019).

\bibitem{Buckley:17}
S.~Buckley \emph{et~al.}, Applied Physics Letters \textbf{111}, 141101 (2017).

\bibitem{Sumikura:14}
H.~Sumikura \emph{et~al.}, Scientific Reports \textbf{4}, 5040 (2014).

\end{thebibliography}

% copied from .bbl file

\end{document}